\documentclass[10pt,conference]{IEEEtran}

\usepackage{amsmath,amssymb}
\usepackage{graphicx}
\usepackage{booktabs}
\usepackage{array}
\usepackage{tabularx}
\usepackage{multirow}
\usepackage{cite}
\usepackage{url}
\usepackage{microtype}
\usepackage{placeins}
\usepackage{xspace}
\usepackage{pifont}
\usepackage{algorithm}
\usepackage{algpseudocode}

\usepackage{todonotes}
\usepackage{xcolor}

\newif\ifshowtodos
\showtodostrue

\graphicspath{{figures/}}
\newcommand{\sys}{\textnormal{PatchKV}\xspace}
\newcommand{\tightfigcaption}{\vspace{-10pt}}
\newcommand{\tightdatacaption}{\vspace{-4pt}}
\newcolumntype{Y}{>{\raggedright\arraybackslash}X}
\newif\ifshowauthors
\showauthorstrue
\ifshowauthors
\IEEEoverridecommandlockouts
\fi
\renewcommand{\arraystretch}{1.05}
\begin{document}

\title{\sys: Efficient KV Cache Recovery for Dynamically Edited LLM Contexts}

\ifshowauthors
\author{
\IEEEauthorblockN{Guotao Yang, Rui Guo, Siwei He, Sheng Chen,
Yitao Hu\IEEEauthorrefmark{1}, and Keqiu Li}
\IEEEauthorblockA{Tianjin University, China}
\thanks{\IEEEauthorrefmark{1}Corresponding author: Yitao Hu
(e-mail: \texttt{yitao@tju.edu.cn}).}
}
\else
\author{}
\fi

\maketitle

\begin{abstract}
Long-running LLM agent workflows often revise interior context spans while retaining long suffixes. Although suffix tokens remain unchanged, altered causal histories and rotary positions prevent exact reuse of their offloaded key--value (KV) states. Full suffix recomputation wastes prefill work, while indiscriminate reuse propagates stale states and full-precision restoration adds data movement. We present PatchKV, a profile-guided recovery system for suffix-preserving revisions. PatchKV decomposes adjacent context versions into an exact prefix, an updated span, and an aligned suffix. It predicts an edit-local dirty region using an offline length-conditioned drift model, augments this region with sparse nonlocal blocks selected from stored attention, and block-rounds their union into a fixed repair set. The remaining suffix blocks are restored from CPU memory using frozen per-block precision tags and a fused path for dequantization, RoPE correction, and KV-page placement. Across three models and three long-context question-answering workloads, PatchKV achieves a $2.51$--$3.85\times$ speedup in mean resume time-to-first-token over full suffix recomputation and a $1.26$--$2.06\times$ speedup over CacheBlend, while matching or exceeding CacheBlend's F1 score in six of nine settings and remaining within 1.36 points in the others.
\end{abstract}


\section{Introduction}

Large language model (LLM) services increasingly execute long-lived, stateful workloads such as agents, coding assistants, retrieval-augmented generation (RAG), and multi-turn conversations~\cite{schick2023toolformer,yao2023react,wu2023autogen,lewis2020rag}. Their prompts are not independent inputs; they are successive versions of an evolving execution state. Between model invocations, an agent may wait seconds or minutes for tools, retrieval, or external services, during which its inactive key--value (KV) cache can be offloaded to CPU memory or a remote tier to release scarce GPU capacity~\cite{sheng2023flexgen,qin2024mooncake}. Before the agent resumes, the controller may replace a tool result, refresh retrieved evidence, revise a plan, or compress earlier history~\cite{jiang2023llmlingua}. Such updates commonly modify an interior context span while preserving both a long prefix and a long textual suffix; we refer to this operation as a \emph{middle edit}.

Middle edits break the exact-match assumption underlying conventional KV reuse. Although the retained suffix preserves its token IDs, each suffix token now observes a different causal prefix and may occupy a different rotary position. Its cached state can therefore diverge from the state produced by an exact prefill of the revised context~\cite{vaswani2017attention,su2021roformer}. Recomputing the entire suffix restores exact dependencies but discards substantial prior work. Directly reusing all suffix KV avoids this computation but can propagate stale state into generation. When the checkpoint is off-GPU, even valid reused state must additionally be fetched and reconstructed before decode. The central challenge is thus not to maximize cache hits, but to \emph{reconstruct only the states that materially affect generation and restore the remaining states at the lowest possible movement cost}.

\begin{figure}[!t]
\centering
\includegraphics[width=\columnwidth]{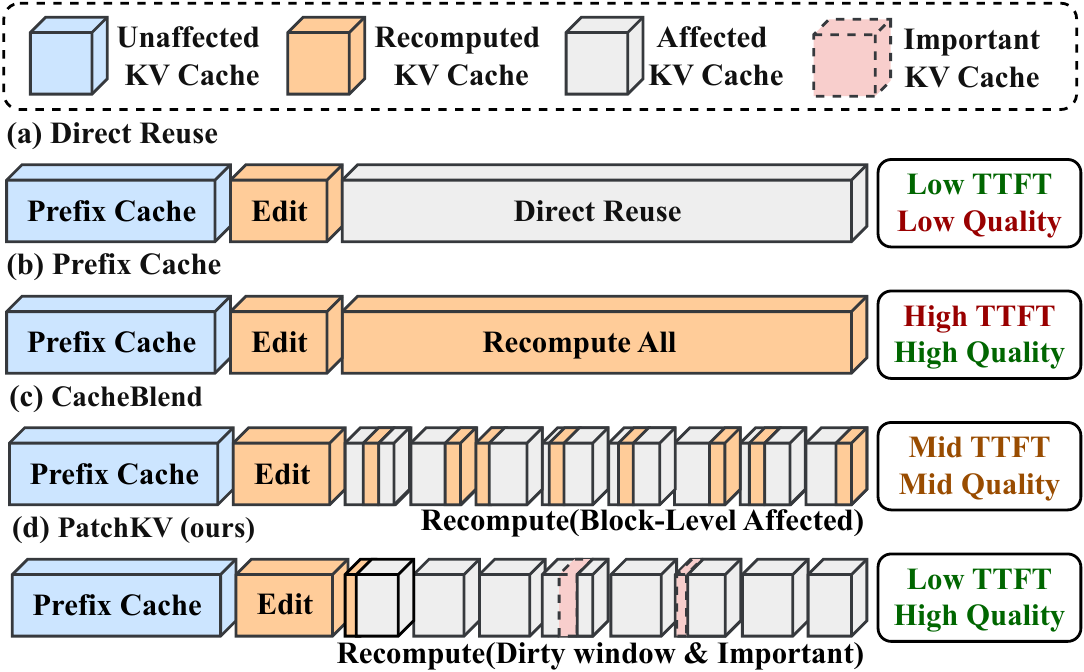}
\tightfigcaption
\caption{Recomputation coverage after a dynamic context update.}
\label{fig:introduction-method-comparison}
\end{figure}

Existing mechanisms address only parts of this challenge, as illustrated by their recomputation scopes in Fig.~\ref{fig:introduction-method-comparison}. Direct KV reuse minimizes computation but retains suffix states affected by the edit. Prefix and radix caches terminate reuse at the first mismatch and consequently prefill the complete suffix~\cite{kwon2023vllm,zheng2024sglang}. Prompt Cache reuses statically declared modules, while CacheBlend composes cached chunks and selectively recomputes high-deviation tokens across them to recover cross-chunk dependencies~\cite{gim2024promptcache,yao2025cacheblend}. For a long suffix inherited from a revised context, direct reuse can preserve stale state, full suffix prefill repeats excessive work, and distributed token repair can recompute more states than those that are both invalid and consequential to generation. KV quantization and streaming reduce storage or transfer cost only after a system has decided which states are semantically reusable~\cite{liu2024kivi,hooper2024kvquant,liu2024cachegen}. None of these approaches jointly determines where edit-induced drift propagates, which distant states remain important, and how the reusable remainder should be represented. \sys concentrates exact reconstruction on the edit-affected region and sparse nonlocal high-impact blocks, then restores the remaining suffix at calibrated precision.

To address these limitations, we derive three observations from middle-edit traces and controlled interventions. First, edit-induced value-cache deviation is concentrated near the edited span and generally decays with suffix distance, while the transition to stable states varies across models. Second, state deviation alone does not identify every important token: distant instructions, query terms, or retrieved facts can remain consequential when they receive high attention. Third, reusable blocks differ in their tolerance to reduced precision. These observations separate the recovery problem into local validity, nonlocal influence, and representation cost.

These observations define the recovery order in \sys. Offline, paired exact prefills train a length-conditioned predictor for the edit-local dirty boundary and calibrate per-block precision tags from reconstruction risk. During checkpoint creation for $C_{\mathrm{old}}$, \sys also records compact attention statistics. At resume time, the predictor marks the local dirty region, attention-guided selection adds sparse nonlocal blocks, and their block-aligned union forms the repair set. \sys recomputes this set under the revised context while restoring the remaining blocks at their stored precisions through a fused path that combines dequantization, RoPE correction, and KV-page placement. Fixing semantic repair before representation selection prevents low-cost transport from masking stale-state errors.

We make three contributions:
\begin{itemize}
    \item We characterize middle-edit KV recovery and identify three structures that make selective recovery tractable: distance-structured state drift, sparse nonlocal generation impact, and heterogeneous precision tolerance among reusable blocks.
    \item We design \sys, which combines length-conditioned drift profiling, attention-guided selective recomputation, and quality-constrained quantized transport into a block-aligned recovery policy for offloaded dynamic contexts.
    \item We implement a CPU-offload research prototype with fused quantization and restoration kernels and evaluate it across three model families and three long-context workloads, observing up to a $3.85\times$ TTFT speedup over Prefix Cache and F1 scores that match or exceed CacheBlend in six of nine settings.
\end{itemize}

\section{Background}
\label{sec:background}

This section characterizes the long-lived agent workloads targeted by this work, defines the middle-edit context model, and explains why recovering an aligned suffix incurs both semantic and data-movement costs.

\subsection{LLM Agents and Dynamic Contexts}

LLM agents interleave inference with tool calls, planning, retrieval, memory management, and communication with other agents~\cite{schick2023toolformer,yao2023react,wu2023autogen}. Their prompts are snapshots of a long-lived execution state: system instructions, retrieved evidence, tool traces, intermediate plans, code state, and prior interactions accumulate as an episode progresses. Long contexts are therefore a routine operating regime for these workloads. Moreover, agent controllers revise these contexts while execution is in progress. They may replace a stale tool result, refresh retrieved evidence, revise a failed plan, or compact earlier history into a summary. RAG and prompt compression create the same update pattern~\cite{lewis2020rag,jiang2023llmlingua}.

\begin{figure}[!t]
\centering
\includegraphics[width=\columnwidth]{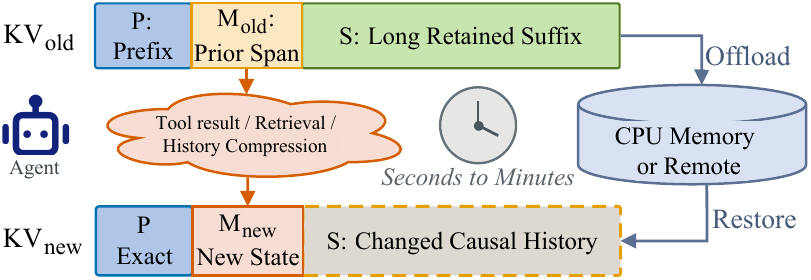}
\tightfigcaption
\caption{Dynamic contexts in long-running agent workloads, where tool execution and context updates create successive prompt versions while KV state may be offloaded.}
\label{fig:agent-dynamic-context}
\end{figure}

Many such updates preserve most of the preceding context and a long textual suffix while changing an interior span. We call the broader evolving-prompt setting a \emph{dynamic context} and this operation a \emph{middle edit}. Its reuse opportunity differs from independent requests or statically composed prompt fragments: the reusable state is inherited from adjacent versions of the same execution, but a potentially long retained suffix has observed a different causal history. Version alignment permits recovery without request-time comparison of old and revised KV tensors or reconstruction of the entire suffix. Fig.~\ref{fig:agent-dynamic-context} illustrates this setting.

We model two adjacent context versions as
\begin{equation}
C_{\mathrm{old}}=P\oplus M_{\mathrm{old}}\oplus S,\qquad
C_{\mathrm{new}}=P\oplus M_{\mathrm{new}}\oplus S,
\label{eq:context}
\end{equation}
where $P$ is the longest common prefix, $S$ is the maximal aligned token-identical suffix, and $M_{\mathrm{old}}$ and $M_{\mathrm{new}}$ are the unmatched spans. This decomposition exposes three distinct state classes. Tokens in $P$ retain both position and causal history and are exact-prefix hits. Tokens in $X=\operatorname{Tokens}(M_{\mathrm{new}})$ have no prior KV states, so those states must be computed. Tokens in $S$ preserve their identities but not necessarily their positions or causal histories, making them candidates for recovery rather than exact cache hits.

\subsection{KV Cache Reuse under Dynamic Updates}

Transformer serving materializes one key and one value per layer and token during prefill and reads those states during autoregressive decode~\cite{vaswani2017attention,yu2022orca}. At layer $l$, the content-dependent state and its rotary-positioned key are
\begin{equation}
\widetilde K_i^l=W_K^l h_i^{l-1},\qquad
V_i^l=W_V^l h_i^{l-1},\qquad
K_i^l=\mathcal{R}(p_i)\widetilde K_i^l,
\end{equation}
where $h_i^{l-1}$ depends on the causal history through token $i$, and $\mathcal{R}(p_i)$ is the rotary position embedding (RoPE) transform at position $p_i$~\cite{su2021roformer}. Replacing $M_{\mathrm{old}}$ can therefore change both $\widetilde K_i^l$ and $V_i^l$ for every aligned suffix token. If the old and new positions are $\omega_i$ and $\rho_i$, respectively, the known phase can be corrected analytically:
\begin{equation}
K_i^{\mathrm{corr},l}=
\mathcal{R}(\rho_i){\mathcal{R}(\omega_i)}^{-1}K_i^{\mathrm{old},l}
=\mathcal{R}(\rho_i)\widetilde K_i^{\mathrm{old},l}.
\label{eq:k-correction}
\end{equation}
This correction changes the stored phase but does not reconstruct the content state $\widetilde K_i^l$ under the revised causal history; values have no analogous position-only correction. Hence token equality and position correction provide an alignment mechanism, not a certificate of KV validity.

Exact prefix caches remain valid for $P$ because its tokens retain the same positions and causal histories~\cite{kwon2023vllm,zheng2024sglang}. The aligned suffix $S$ does not satisfy this condition: even an unchanged suffix token can encode stale dependencies on $M_{\mathrm{old}}$. Modern serving engines further organize KV rows into fixed-size token blocks or pages. Paging does not change this causal property, but it makes a block the practical unit of transfer and reconstruction. A token-level recovery policy must therefore map its decisions onto block-aligned execution without rebuilding the entire long suffix.

\subsection{KV Cache Offloading and Transport}

Agent execution can pause for seconds or minutes while waiting for tools, retrieval, or external services. Retaining the inactive request's KV cache on the GPU during this interval consumes capacity that could serve other requests. A serving system may instead checkpoint the per-layer KV blocks in CPU memory or a remote storage tier and restore them when the agent resumes~\cite{sheng2023flexgen,qin2024mooncake}.

Restoration lies on the resume time-to-first-token (TTFT) path: the serving engine must locate the checkpoint, transfer the required blocks over PCIe or the network, reconstruct their GPU representation, and combine them with any recomputed states before decoding. Low-bit KV formats can reduce payload and transfer cost~\cite{liu2024cachegen,liu2024kivi,hooper2024kvquant}, but representation efficiency and semantic validity are distinct. Quantization changes how a reusable block is moved; it does not establish whether that block remains valid after its causal prefix changes. Middle-edit recovery must therefore first identify which suffix blocks require state reconstruction and then choose an efficient representation for the remainder.

\section{Motivation}
\label{sec:motivation}

We next examine middle-edit KV behavior through three observations. They show that edit-induced state drift is distance-structured, generation impact extends sparsely beyond the local drift region, and reusable blocks differ in precision tolerance.

\subsection{Observation 1: Edit-Induced Value Drift Is Distance-Structured}

Although a middle edit can affect every suffix state, the resulting content-state drift may decay with distance from the edit. We therefore examine whether this drift exhibits a spatial structure that can support localized repair. For an aligned, token-identical suffix token $i$, let $d_i$ denote its token distance from the end of $M_{\mathrm{new}}$ in new-token coordinates:
\begin{equation}
d_i=\operatorname{dist}_{\mathrm{tok}}(i,M_{\mathrm{new}}).
\label{eq:distance}
\end{equation}
Let $V_{\mathrm{old}}^{i,l}$ be the checkpoint value and $V_{\mathrm{full}}^{i,l}$ the value obtained by an exact prefill of $C_{\mathrm{new}}$. Paired old and revised prefills measure the resulting content-state drift as
\begin{equation}
\delta_{V,i}^l=
1-\cos\!\left(V_{\mathrm{old}}^{i,l},V_{\mathrm{full}}^{i,l}\right).
\label{eq:v-distance}
\end{equation}

\begin{figure}[t]
\centering
\includegraphics{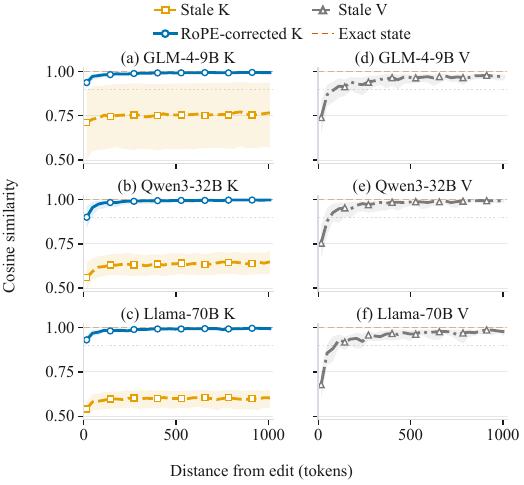}
\tightdatacaption
\caption{KV-state similarity versus suffix distance across three models. Lines are medians and shading is P25--P75 over archived request--layer samples; dashed and dotted guides mark exact-state similarity and 0.90, respectively.}
\label{fig:motivation-measurements}
\end{figure}

Fig.~\ref{fig:motivation-measurements} shows a consistent distance-dependent pattern across all three models. Stale-value similarity rises from $0.68$--$0.75$ near the edit to $0.97$--$0.99$ at the far end of the suffix, indicating that content-state drift is concentrated near the edit and decreases with distance. The transition from unstable to stable states varies across models, suggesting that a fixed distance threshold may not generalize.

Keys also contain position-induced error. RoPE correction raises key similarity to $0.90$--$0.94$ in the first distance block and $0.994$--$0.997$ in the last, showing that it removes most position-induced error but cannot recover content changed by the revised causal history. Position correction therefore removes deterministic key-phase shifts, whereas content-state drift remains in both keys and values.

\noindent\textbf{Takeaway.}
Middle-edit errors are spatially structured but not purely positional: content-state drift decays with suffix distance, while RoPE correction removes only the key-phase error.

\subsection{Observation 2: Attention Reveals Additional Quality-Critical States}

Distance-structured drift identifies where KV states change most, but deviation alone does not determine their influence on generation. Distant instructions, retrieved facts, or query tokens may remain consequential even when they lie beyond the local drift region. We therefore examine whether attention reveals such nonlocal quality-critical states.

To isolate this effect, we fix the drift-based repair set and add an equal budget of blocks selected by attention, distance, or random sampling. Table~\ref{tab:attention-intervention} shows that, across all three models, the highest-attention blocks lie deep in the suffix and improve exact-continuation rate by 10--14 percentage points over the drift-only set; equally sized near-edit, random, and low-attention selections provide no consistent gain. Attention is not a causal certificate~\cite{jain2019attention}, but these results show that it provides a complementary signal for sparse nonlocal influence, consistent with prior observations of persistent attention importance~\cite{zhang2023h2o,liu2023scissorhands,li2024snapkv}.

\begin{table}[t]
\caption{Equal-budget exact-state intervention across three models.}
\label{tab:attention-intervention}
\centering
\footnotesize
\setlength{\tabcolsep}{1.1pt}
\renewcommand{\arraystretch}{0.94}
\begin{tabular}{@{}lccccccc@{}}
\toprule
& \multicolumn{2}{c}{Top-attn. blocks (\%)} & \multicolumn{5}{c}{Exact continuation (\%)} \\
\cmidrule(lr){2-3}\cmidrule(l){4-8}
Model & \shortstack{Attn.\\mass} & Dist. & \shortstack{Drift\\only} & \shortstack{Near\\edit} & Random & \shortstack{Low\\attn.} & \shortstack{Top\\attn.} \\
\midrule
Qwen3-32B     & 25.8 & 83.5 & 81 & 82 & 81 & 82 & 95 \\
GLM-4-9B      & 33.0 & 72.2 & 86 & 86 & 89 & 85 & 97 \\
Llama-3.3-70B & 14.3 & 75.8 & 87 & 87 & 89 & 87 & 97 \\
\bottomrule
\end{tabular}
\end{table}

\noindent\textbf{Takeaway.}
State drift localizes where representations change most, whereas attention reveals a sparse set of distant states whose stale representations remain consequential to generation.

\subsection{Observation 3: Reusable KV Blocks Exhibit Heterogeneous Precision Tolerance}

\begin{figure}[t]
\centering
\includegraphics{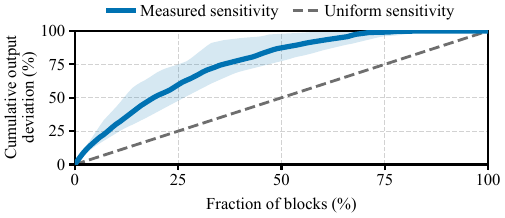}
\tightdatacaption
\caption{Cumulative block-level precision sensitivity across three models.}
\label{fig:precision-sensitivity-concentration}
\end{figure}

Semantic reusability does not imply that every KV block is equally robust to reduced numerical precision. We test this through a controlled blockwise intervention, representing one reusable block at lower precision while keeping every other KV state exact. Fig.~\ref{fig:precision-sensitivity-concentration} shows that, across the three models, the most sensitive 10\% of blocks contribute 30.8\%--42.4\% of the cumulative first-token KL divergence. Thus, precision sensitivity is concentrated in a subset of reusable blocks, motivating block-specific precision selection~\cite{liu2024kivi,hooper2024kvquant}.

\noindent\textbf{Takeaway.}
Reusable KV blocks are not uniformly compressible: precision sensitivity is concentrated in a small subset of blocks.

\section{\sys Design}
\label{sec:design}

\subsection{Overview}

Building on these observations, we present \sys, a unified selective-recovery policy that exploits distance-structured drift, sparse nonlocal influence, and heterogeneous precision tolerance. \sys combines offline KV-drift profiling with online selective KV recovery. Its offline profiler uses paired exact prefills to fit a dirty-window predictor, calibrate model-specific precision thresholds, and assign a frozen precision tag to each block. A CPU-resident KV checkpoint stores clustered suffix blocks together with compact attention and precision metadata. Upon a dynamic context update, the online recovery engine identifies the suffix blocks that require exact recomputation and restores the remaining blocks at their calibrated precisions.

\begin{figure*}[t]
\centering
\includegraphics[width=\textwidth]{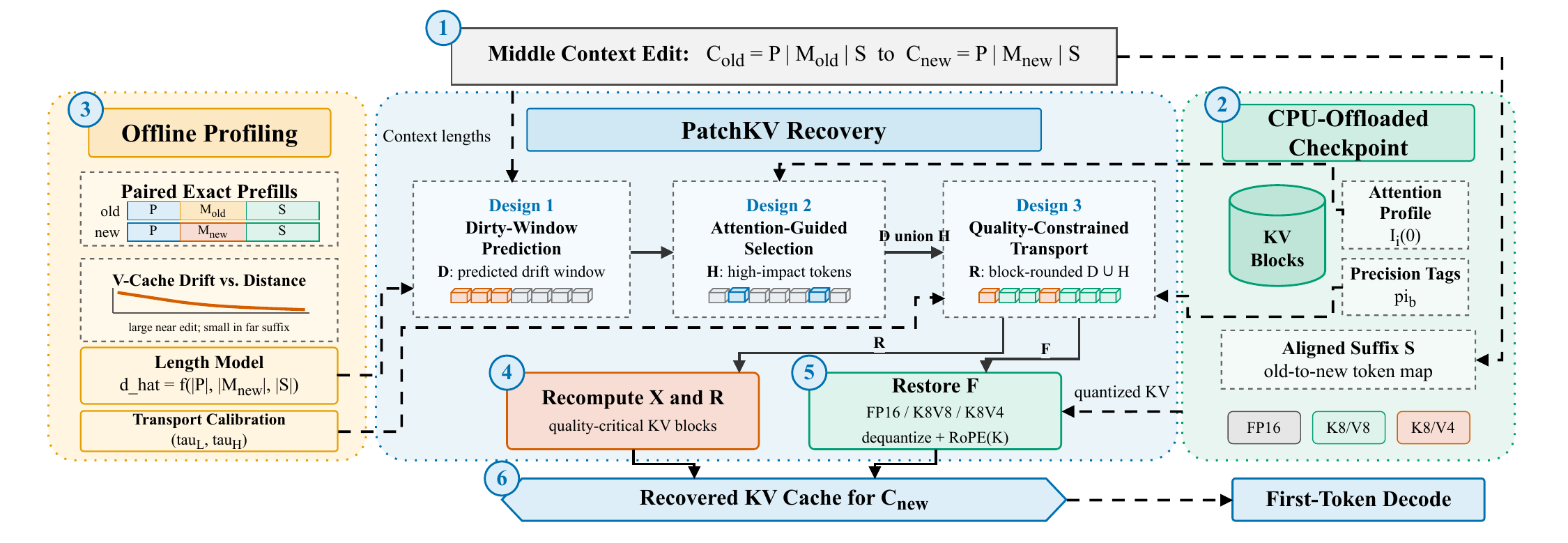}
\tightfigcaption
\caption{Overview of \sys components and recovery workflow.}
\label{fig:patchkv-design-overview}
\end{figure*}

The execution follows the numbered path in Fig.~\ref{fig:patchkv-design-overview}. \ding{172} Context differencing separates the exact prefix, updated span, and aligned suffix, and constructs the old-to-new position mapping. \ding{173} \sys loads the suffix metadata from the CPU-resident checkpoint. \ding{174} The frozen length-conditioned model predicts a local dirty window $D$, while attention-guided selection identifies nonlocal high-impact tokens $H$. Block rounding over $D\cup H$ yields the repair set $R$, and the remaining suffix blocks form the reuse set $F$. \ding{175} \sys exactly recomputes the updated tokens and blocks in $R$ under the new context. \ding{176} It restores blocks in $F$ using their assigned precisions, with fused dequantization, RoPE correction, and page placement. \ding{177} The recomputed and restored blocks are assembled into the KV cache used for first-token decoding.

\subsection{Length-Conditioned Offline Drift Profiling}
\label{sec:design-drift}

\noindent\textit{Insight and Approach.}
After accounting for the deterministic RoPE shift in keys, value-cache deviation measures the content-state changes caused by the revised causal history. This deviation is strongest near the edit and generally decays along the retained suffix. \sys learns a conservative dirty boundary from exact paired prefills and predicts it from context lengths at resume time, avoiding online comparison of old and revised KV states.

During offline profiling, \sys executes exact prefills for paired old and revised contexts, keeping adjacent revisions from the same task in one calibration or validation split. For each aligned suffix block $b$, it measures the layer-aggregated value deviation
\begin{equation}
\delta_{x,b}^V=\operatorname{Agg}_{l\in\mathcal{L}_d,\,i\in b}
\left(1-\cos\!\left(V_{\mathrm{old}}^{i,l},V_{\mathrm{full}}^{i,l}\right)\right),
\label{eq:block-value-drift}
\end{equation}
where $\mathcal{L}_d$ is a small profiled layer set and $\operatorname{Agg}$ is a robust aggregate across layers and tokens. Given a calibrated deviation threshold $\tau_V$ and stability coverage $\gamma$, the oracle boundary for training pair $x$ is
\begin{equation}
d_x^\star=\min\!\left\{d:
\frac{1}{|\mathcal{B}_{x,\ge d}|}
\sum_{b\in\mathcal{B}_{x,\ge d}}\mathbf{1}[\delta_{x,b}^V\le\tau_V]
\ge\gamma\right\},
\label{eq:oracle-dirty-boundary}
\end{equation}
Here, $\mathcal{B}_{x,\ge d}$ contains the suffix blocks in pair $x$ that are at least $d$ tokens from the edit, the minimum is taken only over nonempty tails, and we set $d_x^\star=|S|$ if no candidate distance satisfies the stability criterion.

The resulting labels train a high-quantile length model~\cite{koenker1978regression}
\begin{equation}
\widehat d_x=Q_q\!\left(d_x^\star\mid
\phi(|P|,|M_{\mathrm{new}}|,|S|)\right),
\label{eq:length-conditioned-drift}
\end{equation}
where $\phi$ contains the three log-scaled lengths and their interactions. An upper quantile penalizes underprediction, which can expose stale state, more heavily than moderate overprediction, which only adds recomputation. At resume time, one model evaluation yields
\begin{equation}
D=\{i\in S:d_i<\min(\widehat d_x,|S|)\}.
\label{eq:online-dirty-set}
\end{equation}
New tokens in $X$ remain mandatory because no prior KV exists for them. The runtime artifact for this stage is deliberately compact: the fitted length predictor, its drift calibration, and the execution block size. Attention scoring and stored precision-tag calibration belong to the following stages rather than to the drift profile.

\subsection{Attention-Guided Selective Recomputation}

\noindent\textit{Insight and Approach.}
Value deviation indicates whether a cached state has changed, but not how strongly that state can affect generation. A drift-only policy may therefore miss an important distant instruction or retrieved fact. \sys uses stored attention to rank candidates outside the local dirty set, selects a bounded nonlocal set $H$, and block-rounds $D\cup H$ once before reconstruction.

\begin{figure}[!t]
\centering
\includegraphics[width=\columnwidth]{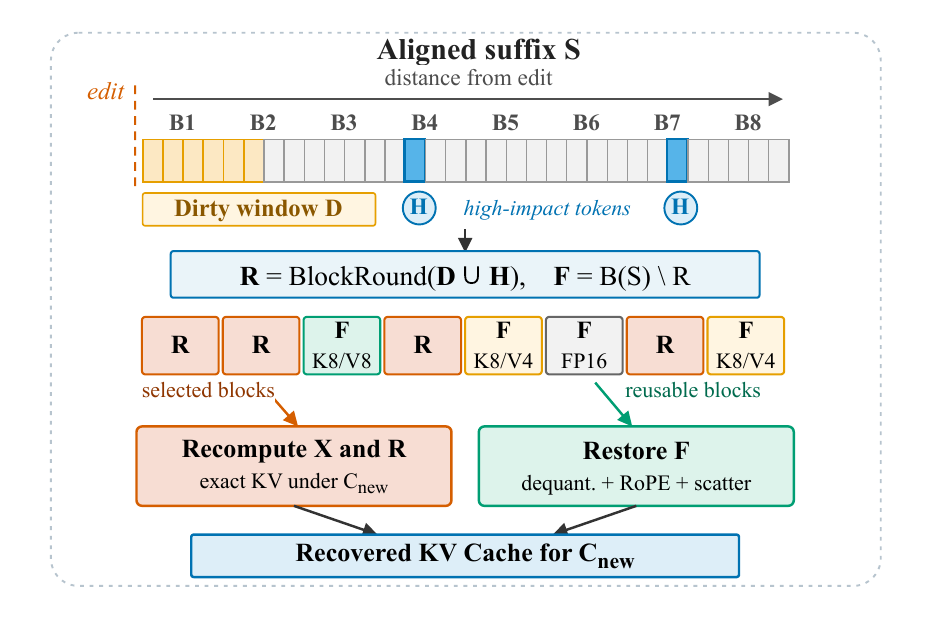}
\tightfigcaption
\caption{Token-to-block partitioning in \sys.}
\label{fig:token-to-block-partition}
\end{figure}

During the $C_{\mathrm{old}}$ checkpoint-creation pass, \sys aggregates attention mass $\bar a_i$ over configured decode query positions, heads, and sampled layers, and stores the normalized score
\begin{equation}
I_i(0)=\operatorname{Norm}(\bar a_i).
\label{eq:importance}
\end{equation}
At resume time, the control plane selects the highest-scoring fraction outside the local dirty set:
\begin{equation}
H=\operatorname{Top}_{\kappa}\{I_i(0):i\in S\setminus D\}.
\label{eq:online-impact-set}
\end{equation}
Let $\mathcal{B}(S)$ denote the blocks covering $S$. For any token set $A\subseteq S$, $\operatorname{BlockRound}_S(A)$ returns the blocks in $\mathcal{B}(S)$ that intersect $A$. The planner fixes the block-aligned repair set and its reusable complement:
\begin{equation}
R=\operatorname{BlockRound}_S(D\cup H),\qquad
F=\mathcal{B}(S)\setminus R.
\label{eq:repair-set}
\end{equation}
This is a semantic partition, not a cost admission rule: every block in $R$ is recomputed under $C_{\mathrm{new}}$, and every block in $F$ is restored from the old cache. Token-aware selection is rounded only for execution so that work aligns with paged-KV blocks, as illustrated in Fig.~\ref{fig:token-to-block-partition}.

The planner computes $D$ and $H$ before launching KV reconstruction, so Eq.~\eqref{eq:repair-set} fixes the request-level block partition $(R,F)$ independently of execution order. The serving engine then recomputes the tokens in $X$ and the blocks in $R$ under $C_{\mathrm{new}}$ and restores every block in $F$ from the checkpoint. For a shifted suffix key, the restore kernel applies Eq.~\eqref{eq:k-correction}; values are never rotated. Decode starts only after both groups have been placed in their destination KV rows.

Stored attention is used as a ranking signal rather than a validity test. Because $R=\operatorname{BlockRound}_S(D\cup H)$ includes every block intersecting $D$ independently of attention ranking, restricting the attention candidates to $S\setminus D$ preserves the local dirty region in the repair set, while the budget $\kappa$ bounds the additional nonlocal recomputation. Drift and attention therefore retain distinct roles: the former identifies unsafe reuse near the edit, and the latter prioritizes potentially influential state beyond that region.

\subsection{Quality-Constrained Quantized Transport}

\noindent\textit{Insight and Approach.}
Semantic reusability does not imply uniform tolerance to low-bit reconstruction error. \sys therefore calibrates a frozen per-block precision policy from attention-weighted reconstruction risk and stores reusable blocks in FP16, K8/V8, or K8/V4 accordingly. Only blocks in $F$ enter this transport stage, where each block is restored directly into its destination KV pages using its frozen precision tag.

For token $i$, let $\widetilde A_i=I_i(0)$ denote its normalized stored attention mass and let $e_i^V$ be the normalized reconstruction residual of a candidate low-bit value representation. \sys forms
\begin{equation}
Z_i=\sqrt{\widetilde A_i e_i^V},\qquad
r_b=\alpha\max_{i\in b}Z_i+(1-\alpha)\operatorname{mean}_{i\in b}Z_i.
\label{eq:transport-risk}
\end{equation}
The maximum term protects a block containing a single fragile token, while the mean term captures block-wide risk. Keys remain at least 8-bit because key error perturbs attention logits; only values enter the 4-bit tier~\cite{liu2024kivi,hooper2024kvquant}. Calibration freezes the block policy
\begin{equation}
\pi_b=\begin{cases}
\mathrm{FP16}, & r_b\ge\tau_H,\\
\mathrm{K8/V8}, & \tau_L\le r_b<\tau_H,\\
\mathrm{K8/V4}, & r_b<\tau_L\land G_4(b),\\
\mathrm{K8/V8}, & r_b<\tau_L\land \neg G_4(b).
\end{cases}
\label{eq:precision-policy}
\end{equation}
Here, $\alpha\in[0,1]$ and $\tau_L<\tau_H$. For blocks with $r_b<\tau_L$, $G_4(b)$ is the calibrated low-bit quality test: passing assigns K8/V4, while failure assigns K8/V8.

The thresholds and the low-bit test are selected on calibration data using measured payload and kernel cost under the configured quality target, then frozen during checkpoint offload. At recovery time, \sys reads the stored $\pi_b$ tags and transports every block in $F$ using its assigned representation.

During offload, a fused \texttt{quantize\_gather} kernel gathers block rows, computes scales, and packs the calibrated K8/V4, K8/V8, or FP16 payloads. Blocks with the same representation are clustered into contiguous runs, so metadata scales with block runs rather than tokens and the payload remains transfer-friendly. During resume, a fused \texttt{dequantize\_scatter} kernel reads the payload, dequantizes it, applies RoPE correction to shifted keys, and writes directly into the request's destination KV pages. The old-to-new token map determines each destination offset when source and destination page boundaries differ.

\section{Evaluation}
\label{sec:evaluation}

\subsection{Experiment Setup}
\label{sec:experiment_setup}

\emph{1) Models and Datasets:}
We evaluate three instruction-tuned models: Qwen3-32B, GLM-4-9B, and Llama-3.3-70B~\cite{yang2025qwen3,glmteam2024chatglm,grattafiori2024llama3}. They range in size from 9B to 70B parameters and differ in layer count, KV-head configuration, and RoPE implementation, covering the model-dependent variation in KV drift, attention concentration, quantization sensitivity, and the cost ratio between recomputation and CPU fetch. We use the same \sys architecture for all models, while freezing model-specific block sizes, attention fractions, and transport precision policies on disjoint calibration requests before formal evaluation. We test HotpotQA-E, 2WikiMQA-E, and TriviaQA-E~\cite{bai2024longbench,yang2018hotpotqa,ho2020twowikimultihopqa,joshi2017triviaqa}. These long-context workloads cover multi-hop, cross-document, and open-domain question answering, respectively, and test whether answer-relevant information survives middle edits amid substantial irrelevant context.

\emph{2) Hardware Environment:}
We run all experiments on one server with 8$\times$ NVIDIA A800 GPUs, two Intel Xeon Gold 5320 CPUs, and 503~GiB of host memory. CPU DRAM is the only KV offload tier. Measured host-to-device bandwidth is 20.4--21.0~GiB/s for pageable memory and 23.2--23.3~GiB/s for pinned memory.

\emph{3) Baseline Methods:}
We compare \sys with three representative recovery strategies:
\begin{itemize}
\setlength{\itemsep}{0pt}
\setlength{\parsep}{0pt}
\setlength{\topsep}{2pt}
\setlength{\partopsep}{0pt}
\item \emph{Prefix Cache} denotes the archived full-suffix-recompute reference, which reuses the exact prefix and rebuilds the complete revised suffix; it therefore provides an exact-state but compute-intensive comparison.
\item \emph{Direct KV Reuse} recomputes the mandatory tokens in the revised span and any block-boundary tokens required for aligned execution, then fetches all remaining suffix KV from CPU DRAM in FP16, exposing both stale-state quality loss and uncompressed movement cost.
\item \emph{CacheBlend} starts from the same stale full-context KV, uses check-layer value-cache drift to select additional tokens for layer-wise recomputation, and fetches the remaining KV in FP16~\cite{yao2025cacheblend}.
\end{itemize}

\emph{4) Implementation Details:}
We implement \sys as a Python research prototype using PyTorch 2.7~\cite{paszke2019pytorch}, Hugging Face Transformers 5.3~\cite{wolf2020transformers}, and Accelerate 1.14~\cite{huggingface2026accelerate}. The prototype captures per-layer KV tensors and attention statistics, constructs block-aligned recovery plans, applies model-aware RoPE correction, and restores reusable KV from CPU memory. CUDA streams and nonblocking memory copies implement asynchronous host-to-device transfer, while CUDA kernels implement quantize-and-gather, INT4 packing, and fused dequantize-and-scatter. KV payload construction, attention profiling, and quantization execute before resume, during user think time or tool execution.

\emph{5) Metrics:}
We use task F1 as the quality metric and mean resume time-to-first-token (TTFT) as the primary latency metric. Resume TTFT is measured from the start of KV recovery execution, after repair-set construction, until the first output token and includes CPU fetch and dequantization, RoPE correction, layer-wise recomputation, and first-token decoding; repair-set construction and work completed before recovery execution are reported separately. We additionally report recomputation ratio, CPU payload, fetch-and-dequantization latency, attention-profiling overhead, precision-planning overhead, and quantize-and-store overhead. All results come from real model executions and measurements using local CPU DRAM offload.

\begin{figure*}[!t]
\centering
\includegraphics{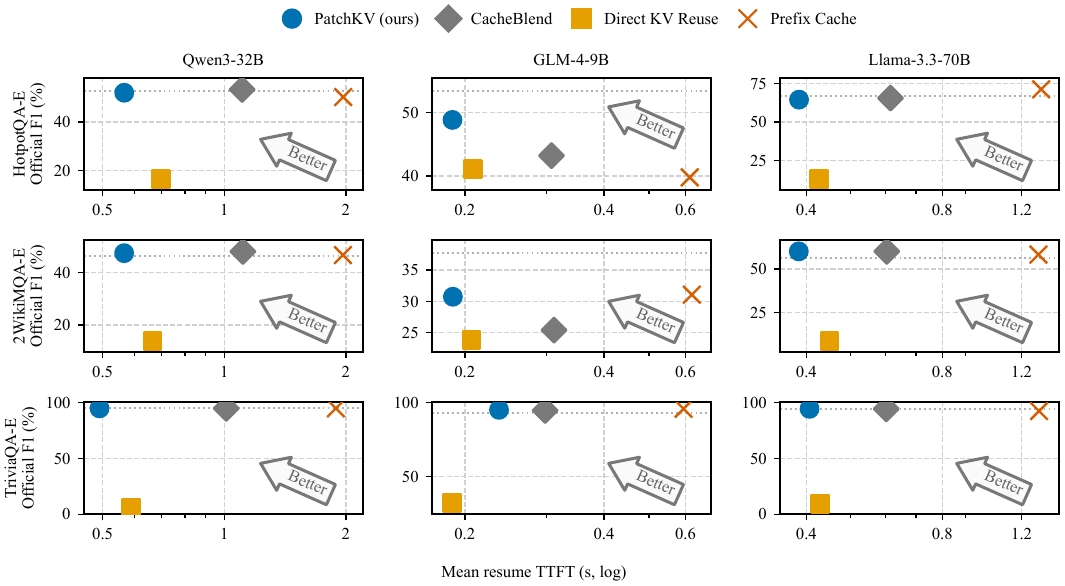}
\tightdatacaption
\caption{End-to-end TTFT and F1 score across three models and three long-context QA workloads.}
\label{fig:evaluation-e2e}
\end{figure*}

\subsection{End-to-End Latency and Quality}
\label{sec:evaluation-e2e}

Fig.~\ref{fig:evaluation-e2e} compares \sys with Prefix Cache, Direct KV Reuse, and CacheBlend over nine model--dataset combinations. Better operating points have lower resume TTFT and higher F1 score.

Across the nine cells, \sys is $2.51$--$3.85\times$ faster than Prefix Cache (median $3.40\times$). It matches or improves the Prefix Cache F1 score on all Qwen3-32B tasks and on two Llama-3.3-70B tasks. On GLM-4-9B, two quality gaps remain within 0.67 points, while the HotpotQA-E score improves by 9.0 points. Across the three Llama-3.3-70B tasks, \sys recovers 51.5--85.1 F1 score points over Direct KV Reuse while further reducing mean resume TTFT by 5.2\%--14.5\%, demonstrating that selective repair avoids both stale-state quality loss and full-precision KV movement. In several cells, \sys also exceeds Prefix Cache in F1 score. Context compression is lossy: Prefix Cache rebuilds suffix states under the summarized context, whereas \sys retains stable KV states formed under the richer pre-compression history. Reusing these states can preserve evidence omitted by the summary and thereby improve answer quality.

\sys is also $1.26$--$2.06\times$ faster than CacheBlend; its F1 score is higher in three cells, tied in three, and at most 1.36 points lower in the remaining cells. Against Direct KV Reuse, it improves F1 score by 6.85--88.99 points and lowers TTFT in eight cells by reducing the transferred payload through lower-precision block representations. Semantic repair and transport reduction must therefore be optimized jointly.

The comparisons separate the sources of the end-to-end gain. Relative to Prefix Cache, the main opportunity is avoiding reconstruction of most aligned suffix blocks. Relative to Direct KV Reuse, the quality recovery comes from repairing locally drifted and nonlocal important state, while the latency reduction comes from lowering the representation cost of the remainder. CacheBlend already performs selective recomputation, so the remaining difference reflects the combined effect of edit-local repair and quantized restoration. The variation across models and datasets also supports model-specific offline calibration rather than one recovery ratio or precision policy shared by every deployment.

\begin{figure*}[t]
\centering
\includegraphics{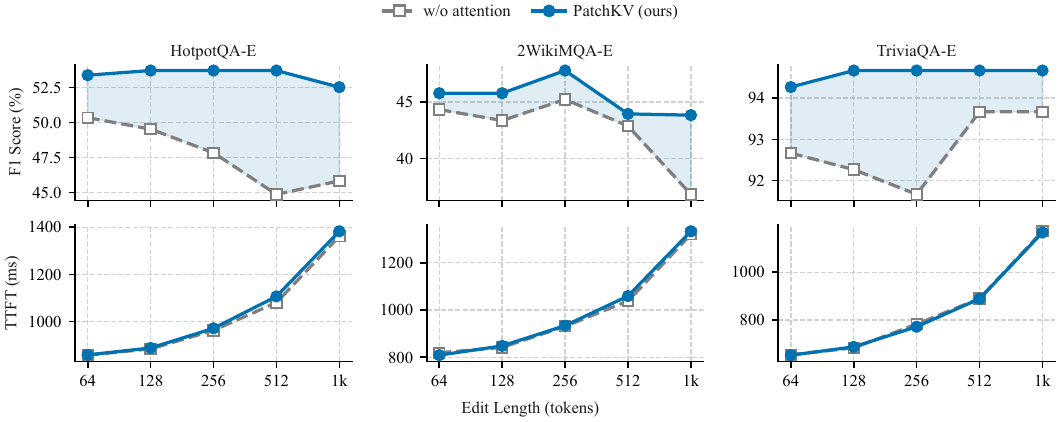}
\tightdatacaption
\caption{Attention-guided selective recomputation ablation on Qwen3-32B.}
\label{fig:evaluation-attention-selection}
\end{figure*}

\subsection{Design Ablation}
\label{sec:evaluation-ablation}

The following ablations isolate the three decisions in \sys: length-conditioned offline drift profiling, attention-guided selective recomputation, and quality-constrained quantized transport.

\subsubsection{Length-Conditioned Offline Drift Profiling}
We fix attention selection, block size, and transport, and vary only the edit-local dirty-window length.

As shown in Table~\ref{tab:evaluation-dirty-window}, the profiler-selected dirty window has mean F1 scores that are 2.0, 1.0, and 2.0 points higher on HotpotQA-E, 2WikiMQA-E, and TriviaQA-E, at incremental TTFT costs of 24.1, 35.1, and 5.4~ms. Expanding the recomputation range from 64 to 1,024 tokens provides no consistent mean F1 improvement and adds $380.9$--$503.8~ms$ of TTFT across the three workloads. These mean results identify the selected window as a compact latency--quality operating point, while the paired F1 intervals for all nontrivial comparisons include zero.

\subsubsection{Attention-Guided Selective Recomputation}
To isolate attention-guided selective recomputation, we compare the dirty window alone with the same window augmented by the top-scoring 1.5\% of suffix tokens by attention, followed by block rounding, across five edit lengths.

Fig.~\ref{fig:evaluation-attention-selection} shows that attention-guided repair improves mean F1 score in every evaluated edit condition, with gains of 1.0--8.86 points (3.48 on average). The actual recomputation ratio increases by 0.29--0.60 percentage points, while mean TTFT increases by 5.3~ms on average. Paired bootstrap intervals~\cite{efron1993bootstrap} are nonnegative in most settings and strictly positive in several, supporting the positive mean trend. The dirty window repairs locally unstable state, while attention selection recovers sparse distant state that remains consequential to generation.

\subsubsection{Quality-Constrained Quantized Transport}
We hold the repair set fixed and vary only the KV representation; all four methods use the same fused RoPE-remap kernel, pinned host payload, and nonblocking transfer path.

Relative to FP16, \sys reduces mean resume TTFT by 15.0\% with a 0.35-point F1 score reduction. It is 5.6\% faster than INT8 while its F1 score remains within 0.05 points of INT8. INT4 is 3.5--3.7\% faster than \sys but loses 1.15 F1 score points. No uniform precision therefore provides both the observed quality of INT8 and the latency of the mixed policy across all three workloads.

\begin{table}[t]
\caption{Effect of the selected 64-token dirty window on Qwen3-32B.}
\label{tab:evaluation-dirty-window}
\centering
\footnotesize
\setlength{\tabcolsep}{2.8pt}
\renewcommand{\arraystretch}{0.94}
\begin{tabular}{@{}lcccc@{}}
\toprule
Dataset & Metric & 0 tokens & 64 tokens & Change \\
\midrule
\multirow{2}{*}{HotpotQA-E} & F1 score & 51.72 & 53.72 & +2.00 \\
            & TTFT (ms) & 618.92 & 643.01 & +24.09 \\
\addlinespace[1pt]
\multirow{2}{*}{2WikiMQA-E} & F1 score & 45.10 & 46.10 & +1.00 \\
            & TTFT (ms) & 621.30 & 656.37 & +35.07 \\
\addlinespace[1pt]
\multirow{2}{*}{TriviaQA-E} & F1 score & 92.67 & 94.67 & +2.00 \\
            & TTFT (ms) & 587.62 & 593.06 & +5.44 \\
\bottomrule
\end{tabular}
\end{table}

\begin{figure}[t]
\centering
\includegraphics{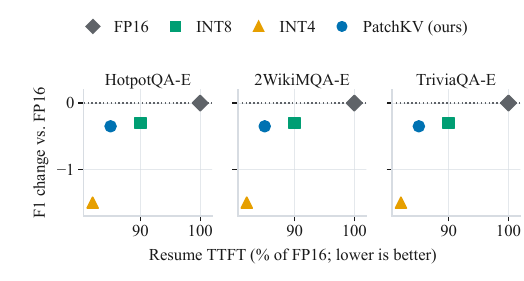}
\tightdatacaption
\caption{Quantized-transport ablation on Llama-3.3-70B. TTFT is normalized to FP16, and F1 score is shown as the difference from FP16.}
\label{fig:evaluation-quantized-transport}
\end{figure}

\subsection{Sensitivity Analysis}
\label{sec:evaluation-sensitivity}

We vary the actual recomputation ratio after block rounding and the KV block size used for selection and recovery.

\begin{figure*}[!t]
\centering
\includegraphics{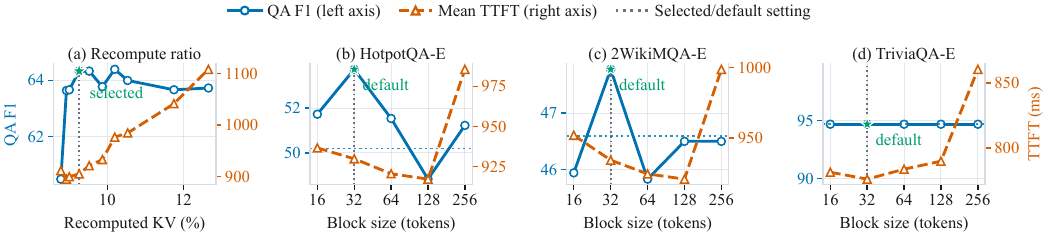}
\tightdatacaption
\caption{Sensitivity analysis on Qwen3-32B.}
\label{fig:evaluation-sensitivity}
\end{figure*}

Fig.~\ref{fig:evaluation-sensitivity}(a) reports the repair set after dirty-window union and block expansion. Raising the attention fraction from 0 to 1.5\% increases the actual ratio from 8.78\% to 9.26\% and F1 score from 60.47 to 64.32, at 905.4~ms TTFT. A 2\% fraction reaches the same F1 score but costs 921.4~ms; larger ratios increase TTFT to 1,107.5~ms without monotonic quality gains. No evaluated attention fraction stays within the predefined 0.25-point quality tolerance of the best setting on every dataset. Calibration therefore selects 1.5\% by the predefined minimax rule, which first minimizes the worst per-dataset quality gap and then minimizes mean TTFT.

A 32-token block size achieves the highest mean F1 score on HotpotQA-E and 2WikiMQA-E and, among quality-equivalent TriviaQA-E points, the lowest TTFT (Fig.~\ref{fig:evaluation-sensitivity}(b)--(d)). A 128-token block size is 12.7--13.5~ms faster on the first two tasks but has mean F1 scores that are 4.89 and 1.27 points lower. On an independent request set, 32-token blocks also have mean F1 scores that are 1.22--2.19 points higher than those of 64-token blocks while reducing TTFT by 22.4--38.2~ms. We therefore use 32-token blocks by default.

\subsection{Overhead Analysis}
\label{sec:evaluation-overhead}

\textbf{Attention profiling.}
On Qwen3-32B with 10k-token contexts, collecting attention statistics takes 122.9--125.2~ms across the three workloads. Token-to-block aggregation adds 1.36--4.74~ms, and constructing the repair set adds 2.62--3.30~ms. The attention-forward and aggregation path accounts for 3.47--3.58\% of the original prefill time. Statistics are collected before offload, leaving only about 3~ms of resume-time set construction.

\textbf{Quantization and restoration.}
In a five-repeat Qwen3-32B CPU DRAM microbenchmark with an 8k-token KV checkpoint, FP16, INT8, INT4, and \sys require 2,007.8, 1,023.4, 523.4, and 799.4~MiB of CPU memory, respectively. Their measured fetch-and-dequantization times are 344.4, 183.6, 107.2, and 158.0~ms. Thus, \sys reduces payload by 21.9\% and restoration time by 13.9\% relative to INT8, while reducing both by 60.2\% and 54.1\% relative to FP16. Its 4.44-ms precision-planning step and 336.3-ms quantize-and-store step occur during offload and are outside resume TTFT.

\section{Related Work}
\label{sec:related-work}

\subsection{Prompt and KV Cache Reuse}

Modern LLM serving systems reuse previously computed KV state to reduce prefill cost. vLLM's PagedAttention improves KV-memory utilization through paging, while SGLang's radix cache shares common prompt prefixes across requests~\cite{kwon2023vllm,zheng2024sglang}. SGLang's exact prefix reuse terminates at the first token mismatch; after a middle edit, the retained suffix must be recomputed. Prompt Cache instead exposes reusable prompt modules, but assumes that module boundaries and composition are declared in advance~\cite{gim2024promptcache}. CacheBlend composes cached knowledge chunks for RAG and repairs missing cross-chunk dependencies by distributing partial recomputation across the reused chunks~\cite{yao2025cacheblend}. \sys targets a different reuse pattern: two adjacent versions of a long-lived context share an aligned textual suffix whose cached state is no longer exact. It exploits the spatial structure of edit-induced drift to localize repair, then adds sparse nonlocal blocks whose state remains important to generation.

\subsection{KV Cache Compression and Offloading}

KV offloading and disaggregated inference move inactive or phase-separated state away from scarce accelerator memory. FlexGen offloads model and runtime state across GPU, CPU, and storage, while Mooncake organizes serving around a disaggregated KV-cache tier~\cite{sheng2023flexgen,qin2024mooncake}. DistServe and Splitwise separate prefill and decode resources to improve serving goodput~\cite{zhong2024distserve,patel2024splitwise}. Complementary work reduces the representation cost of KV state: CacheGen compresses KV tensors for streaming, and KIVI and KVQuant develop low-bit KV formats for efficient long-context inference~\cite{liu2024cachegen,liu2024kivi,hooper2024kvquant}. These systems reduce residency or encoding cost without testing whether a suffix remains semantically valid after its causal prefix changes. \sys fixes the repair set for the revised context before applying mixed precision to the reusable complement.

\subsection{Attention-Guided KV Selection}

Several methods exploit nonuniform token importance to reduce KV-cache capacity. H$_2$O retains attention heavy hitters, Scissorhands builds on the persistence of token importance, and SnapKV selects prompt positions from an observation window before generation~\cite{zhang2023h2o,liu2023scissorhands,li2024snapkv}. Their primary objective is cache eviction or compression within a single causal execution. \sys instead recovers state across two context versions, where attention alone cannot reveal which unchanged suffix states have drifted because of an earlier edit. It uses attention after drift-based localization to select distant blocks whose exact reconstruction can affect generation, keeping state invalidity separate from downstream importance.

\section{Conclusion}
\label{sec:conclusion}

Middle edits can preserve suffix token IDs while changing their causal states and RoPE positions. \sys addresses this mismatch by predicting an edit-local dirty window from offline drift profiles, augmenting it with sparse attention-selected nonlocal blocks, and rounding their union to the serving engine's KV-block boundaries. It exactly recomputes this repair set while restoring the remaining aligned suffix at calibrated per-block precision through a fused path for dequantization, RoPE correction, and page placement. Across the evaluated workloads, the prototype achieves up to a $3.85\times$ TTFT speedup over Prefix Cache and matches or exceeds CacheBlend's F1 score in six of nine settings.

\bibliographystyle{IEEEtran}
\bibliography{references}

\end{document}